\documentclass[2column]{aa} 
\usepackage{natbib}
\usepackage{graphicx}
\usepackage{txfonts}

\begin{document}
 
\title{Formation and Evolution of Planetary Systems in Presence of Highly Inclined Stellar Perturbers}  
\titlerunning{Planet Formation in Inclined Binaries}
\author{Konstantin Batygin\inst{1,2} \and Alessandro Morbidelli\inst{1} \and Kleomenis Tsiganis\inst{3}} 
\authorrunning{Batygin et al.}
\institute{Departement Cassiop$\mathrm{\acute{e}}$e: Universite de Nice-Sophia Antipolis, Observatoire de la C$\mathrm{\hat{o}}$te dÕAzur, 06304 Nice, France
\and
Division of Geological and Planetary Sciences, California Institute of Technology, Pasadena, CA 91125
\and
Section of Astrophysics, Astronomy \& Mechanics, Department of Phyiscs, Aristotle University of Thessaloniki, GR 54 124 Thessaloniki, Greece}
\abstract
{The presence of highly eccentric extrasolar planets in binary stellar systems suggests that the Kozai effect has played an important role in shaping their dynamical architectures. However, the $\textit{formation}$ of planets in inclined binary systems poses a considerable theoretical challenge, as orbital excitation due to the Kozai resonance implies destructive, high-velocity collisions among planetesimals.}
{To resolve the apparent difficulties posed by Kozai resonance, we seek to identify the primary physical processes responsible for inhibiting the action of Kozai cycles in protoplanetary disks. Subsequently, we seek to understand how newly-formed planetary systems transition to their observed, Kozai-dominated dynamical states.}
{The main focus of this study is on understanding the important mechanisms at play. Thus, we rely primarily on analytical perturbation theory in our calculations. Where the analytical approach fails to suffice, we perform numerical $N$-body experiments.}
{We find that theoretical difficulties in planet formation arising from the presence of a distant ($\tilde{a} \sim 1000$AU) companion star, posed by the Kozai effect and other secular perturbations, can be overcome by a proper account of gravitational interactions within the protoplanetary disk. In particular, fast apsidal recession induced by disk self-gravity tends to erase the Kozai effect, and ensure that the disk's unwarped, rigid structure is maintained. Subsequently, once a planetary system has formed, the Kozai effect can continue to be wiped out as a result of apsidal precession, arising from planet-planet interactions. However, if such a system undergoes a dynamical instability, its architecture may change in such a way that the Kozai effect becomes operative.}
{The results presented here suggest that planetary formation in highly inclined binary systems is not stalled by perturbations, arising from the stellar companion. Consequently, planet formation in binary stars is probably no different from that around single stars on a qualitative level. Furthermore, it is likely that systems where the Kozai effect operates, underwent a transient phase of dynamical instability in the past. }
\maketitle

\section{Introduction}
Among the most unexpected discoveries brought forth by a continually growing collection of extra-solar planets has been the realization that giant planets can have near-parabolic orbits. Since the seminal discovery of 16Cygni B \citep{1997ApJ...483..457C}, followed by HD80606 \citep{2001A&A...375L..27N}, much effort has been dedicated to understanding the dynamical origin and evolution of systems with highly eccentric planets. In particular, it has been understood that in presence of a companion star on an inclined orbital plane, the most likely pathway to production of such extreme planet eccentricities is via Kozai resonance \citep{2001ApJ...562.1012E}.

The Kozai resonance was first discovered in the context of orbital dynamics of highly-inclined asteroids forced by Jupiter, and has been subsequently recognized as an important process in sculpting the asteroid belt \citep{1962AJ.....67..591K} as well as being the primary mechanism by which long-period comets become Sun-grazing \citep{1992A&A...257..315B, 1996CeMDA..64..209T}. Physically, the Kozai resonance corresponds to extensive excursions in eccentricity and inclination of a test particle forced by a massive perturber, subject to conservation of the third Delaunay momentum $H = \sqrt{1-e^2} \cos(i)$ (where $e$ is the eccentricity and $i$ is the inclination), and libration of its argument of perihelion $\omega$ around $\pm 90^\circ$. A necessary criterion for the resonance is a sufficiently large inclination ($i > \arccos \sqrt{3/5}$) relative to the massive perturber's orbital plane, during the part of the cycle where the test-particle's orbit is circular.

By direct analogy with the Sun-Jupiter-asteroid picture, the Kozai resonance can give rise to variation in orbital eccentricity and inclination of an extra-solar planet, whose orbit, at the time of formation, is inclined with respect to a stellar companion of the
planet's host star \citep{2003ApJ...589..605W}. In the systems mentioned above (16Cygni B, HD80606) the stellar companions' (e.g. 16Cygni A, HD80607) proper motion has been verified to be consistent with a binary solution. Other examples of planets in binary stellar systems are now plentiful (e.g. $\gamma$ Cephei \citep{2003ApJ...599.1383H}, HD 196885 \citep{2008A&A...479..271C}, etc) with binary separation spanning a wide range ($\tilde{a} \sim 10 - 1000$ AU). However, all planets whose eccentricities are expected to have been excited by the Kozai resonance with the companion star are in wide binaries.

If a Kozai cycle is characterized by a sufficiently small perihelion distance, the eccentricity of the planet may subsequently decay tidally, yielding a pathway to production of hot Jupiters, whose orbital angular momentum vector is mis-aligned with respect to the stellar rotation axis \citep{2007ApJ...669.1298F}. The presence of such objects has been confirmed via observations of the Rossiter-McLaughlin effect \citep{1924ApJ....60...22M}, leading to a notion that Kozai cycles with tidal friction are responsible for
generating at least some misaligned systems \citep{2010ApJ...718L.145W, 2011ApJ...729..138M}.

Kozai cycles may have also played an important role in systems where a stellar companion is not currently observed. Indeed, one can envision an evolutionary history where the binary companion gets stripped away as the birth cluster disperses. In fact, such a scenario may be rather likely, as the majority of stars are born in binary systems \citep{1991A&A...248..485D}. In this case, a Kozai cycle can be suddenly interrupted, causing the planet's eccentricity to become "frozen-in."

In face of the observationally suggested importance of Kozai cycles during early epochs of planetary systems' dynamical evolution, the \textit{formation} of planets in presence of a massive, inclined perturber poses a significant theoretical challenge
\citep{1996MNRAS.282..597L, 2009A&A...507..505M, 2010A&A...524A..13T}. After all, in the context of the restricted problem (where only the stars are treated as massive perturbers), one would expect the protoplanetary disk to undergo significant excursions in eccentricity and inclination due to the Kozai resonance, with different temporal phases at different radial distances, resulting in an incoherent structure. Such a disk would be characterized by high-velocity impacts among newly-formed planetesimals, strongly inhibiting formation of more massive objects (planetary embryos) \citep{1993ARA&A..31..129L}.

Damping of eccentricities due to gas-drag has been considered as an orbital stabilization process. However, excitation of mutual inclination among neighboring annuli renders this mechanism ineffective \citep{2009A&A...507..505M}. Ultimately, in the context of a restricted model, one is forced to resort to competing time-scales for formation of planetesimals and dynamical excitation by the companion star. Such an analysis suggests that although possible, planetary formation in binary systems is an unlikely event.

Here, we show that the theoretical difficulties in planet formation arising from the companion star, posed by the Kozai effect and other secular perturbations, can be resolved by a proper account of the self-gravity of the proto-planetary disk (i.e. planetesimals embedded in a gaseous disk). During the preparation of this manuscript, a paper was published \citep{2011A&A...528A..40F} addressing the role of the gravity of a gas disk on the relative motion of embedded planetesimals, with hydrodynamical
simulations. The work of  \cite{2011A&A...528A..40F} neglects the gravitational effects of the gas-disk onto itself, and therefore considers a case where pressure and viscosity keep the disk more coherent against external perturbations than would be possible with self-gravity alone \citep{2010A&A...511A..77F}. This parameter regime is characteristic of systems where external stresses are strong enough to partially overcome the role of self-gravity, but not that of the internal forces of the fluid. Examples of such systems include binary stars with moderate separations ($60$ AU in the simulations of  \cite{2011A&A...528A..40F}).

\begin{figure}[t]
\includegraphics[width=0.5\textwidth]{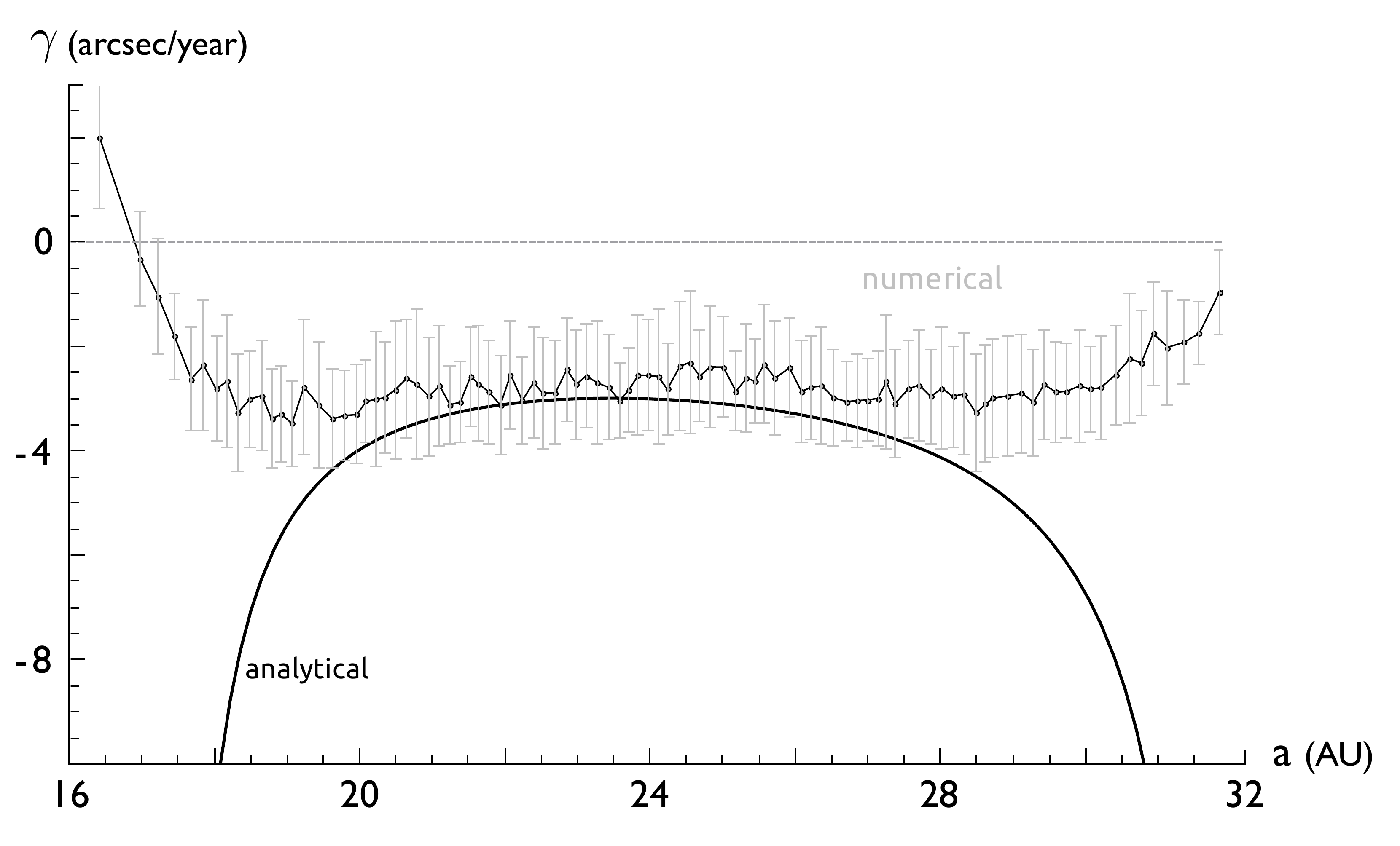}
\caption{An example of apsidal precession,$\gamma$, in a self-gravitating disk. Here the disk is assumed to contain $50 M_{\oplus}$ between 16 and 32 AU, characteristic of a typical post-formation debris disk in the Nice model of solar system formation \citep{2005Natur.435..459T}. The solid curve shows the precession rate predicted by eqs. (1)-(3), as a function of semi major axis. The dots and error bars show the results of a numerical calculation, integrating 3,000 equal-mass particles with a softening parameter of $\epsilon \approx 0.005~$AU to smooth the effects of their mutual close encounters. The disk was binned into 100 annuli in $a$ and the mean frequency of the longitude of pericenter $\dot{\varpi}$ was measured from the time-series of $\varpi$ of the particles in each bin (dots) as well as its variance (error bars). Note that the precession frequency of a self-gravitating disk is negative.}
\label{MENIOS}
\end{figure}

In systems of this sort, the dynamics of the planetesimals tends to be somewhat different from those of the gas. Consequently, gas-drag induces size-sorted orbital evolutions, ultimately leading to high-velocity impacts among planetesimals of different sizes.  This again, limits the prospects for accretion. Conversely, in the present paper we consider binary separation on the order of $\tilde{a} \sim 1000$ AU, consistent with the cases of 16Cyg B and HD HD80606. This allows us to show, with a simple analytic approach, that gravity is a sufficient mechanism to maintain orbital coherence and planetary growth, without any need to account for other forces acting inside the disk. In fact, we show that one of the primary effects of self-gravity is to induce a fast, rigid recession in the longitudes of perihelion and ascending node of the disk.  This allows for planetary formation to take place, as if the secular perturbations arising from the stellar companion were not present. It is noteworthy that such a process is in play, for instance, in the Uranian satellite system, where the Kozai effect arising from the Sun is wiped out owing to secular interactions among the satellites and the precession arising from Uranus' oblateness \citep{2002mcma.book.....M}.

Furthermore, we show that even after the formation process is complete, and the disk has evaporated, the Kozai effect may continue to be wiped out by the orbital precession, arising from planet-planet interactions. This is again in line with the example of the outer solar system, where interactions among the giant planets erase a Kozai-like excitation due to the galactic tide \citep{2007ApJ...669.1298F}. However, if such a planetary system undergoes a dynamical instability, which leads to a considerable
change in system architecture, it may evolve to a state where the Kozai resonance is no-longer inhibited.
 
The purpose of this work is to identify the important physical processes at play, rather than to perform precise numerical simulations. Consequently, we take a primarily analytical approach in addressing the problem. The plan of this paper is as follows. In
section 2, we compute the precession rate, arising from the self-gravity of the disk and show that it is copiously sufficient to impede the Kozai effect. In section 3, we show that under secular perturbations from the companion star, the reference plane of the disk precesses rigidly, implying an un-warped structure. In section 4, we show how an initially stable two-planet system enters the Kozai resonance after a transient instability causes one of the planets to be ejected from the system. We summarize and discuss our results in section 5. 

\section{Kozai Resonance in a Self-Gravitating Disk}

We begin by considering the secular dynamics of planetesimals in an isolated, flat nearly-circular disk of total mass $M_{\mathrm{disk}}$ around a Sun-like ($M_{\star} = 1 M_{\odot}$) star. Due to a nearly-null angular momentum deficit, secular interactions within the disk will not excite the eccentricities and inclinations significantly. Rather, as already mentioned above, the primary effect of disk self-gravity is to induce a fast, retrograde apsidal precession.

Our calculation of the induced precession follows the formalism of \cite{1987gady.book.....B}, originally developed in the context of galactic dynamics. We work in terms of a polar coordinate system, where the radial coordinate is logarithmic ($\rho = \ln r$) and $\phi$ denotes the polar angle. The reduced potential due to a disk surface density $\sigma$ reads:
\begin{equation}
  \Phi = - \frac{{\cal G}}{\sqrt{2}} \int_{-\infty}^{\infty}
  \int_{0}^{2 \pi} \frac{ \mathbf{e}^{\rho/2} \sigma}{\sqrt{ \cosh
  (\rho-\rho') - \cos (\phi - \phi')   }} d\phi' d\rho'\ ,
\end{equation}
where ${\cal G}$ is the gravitational constant. We assume $\sigma \propto r^{-1}$. Consequently, axial symmetry is implicit, and the potential is only a function of $\rho$. The characteristic frequencies of a planetesimal in the disk are the mean motion, $n$, and the radial frequency, $\kappa$:
\begin{eqnarray}
n &=& \frac{1}{a} \left( \frac{\partial \Phi}{\partial r}  \right) \nonumber \\
\kappa &=& \frac{3}{a} \left( \frac{\partial \Phi}{\partial r}  \right) +  \left( \frac{\partial^2 \Phi}{\partial r^2}  \right)
\end{eqnarray}
where $a$ is semi-major axis. The apsidal precession that results from self-gravity, $\gamma$, can then be written as the difference between the mean motion and radial frequencies
\begin{equation}
\gamma \equiv n - \kappa.
\end{equation}

In practice, the calculation of $\gamma$ is performed by breaking up the disk into cells and computing the derivatives discretely. Following \cite{2007Icar..189..196L}, we split the disk into 1000 logarithmic radial annuli, and take the angular cell width to
be $\Delta \phi = 0.5^{\circ}$. We assume the disk edges to be $a_{in} = 0.5$AU and $a_{out} = 50$AU, although the results are not particularly sensitive to these choices. The resulting precession in the disk is roughly uniform in $a$, except for the edges, where this linear theory breaks down. Numerical experiments of debris disks, where self-gravity is taken into account directly, however, show that the precession rate at the edges is also roughly uniform and quantitatively close to that elsewhere in the disk (see Figure 1). In other words, the disk's apsidal precession is approximately rigid. Consequently, for the purposes of this work, we take the precession rate evaluated at $a = 10$AU, to be the characteristic $\gamma$ for the entire disk.

\begin{figure}[t]
\includegraphics[width=0.5\textwidth]{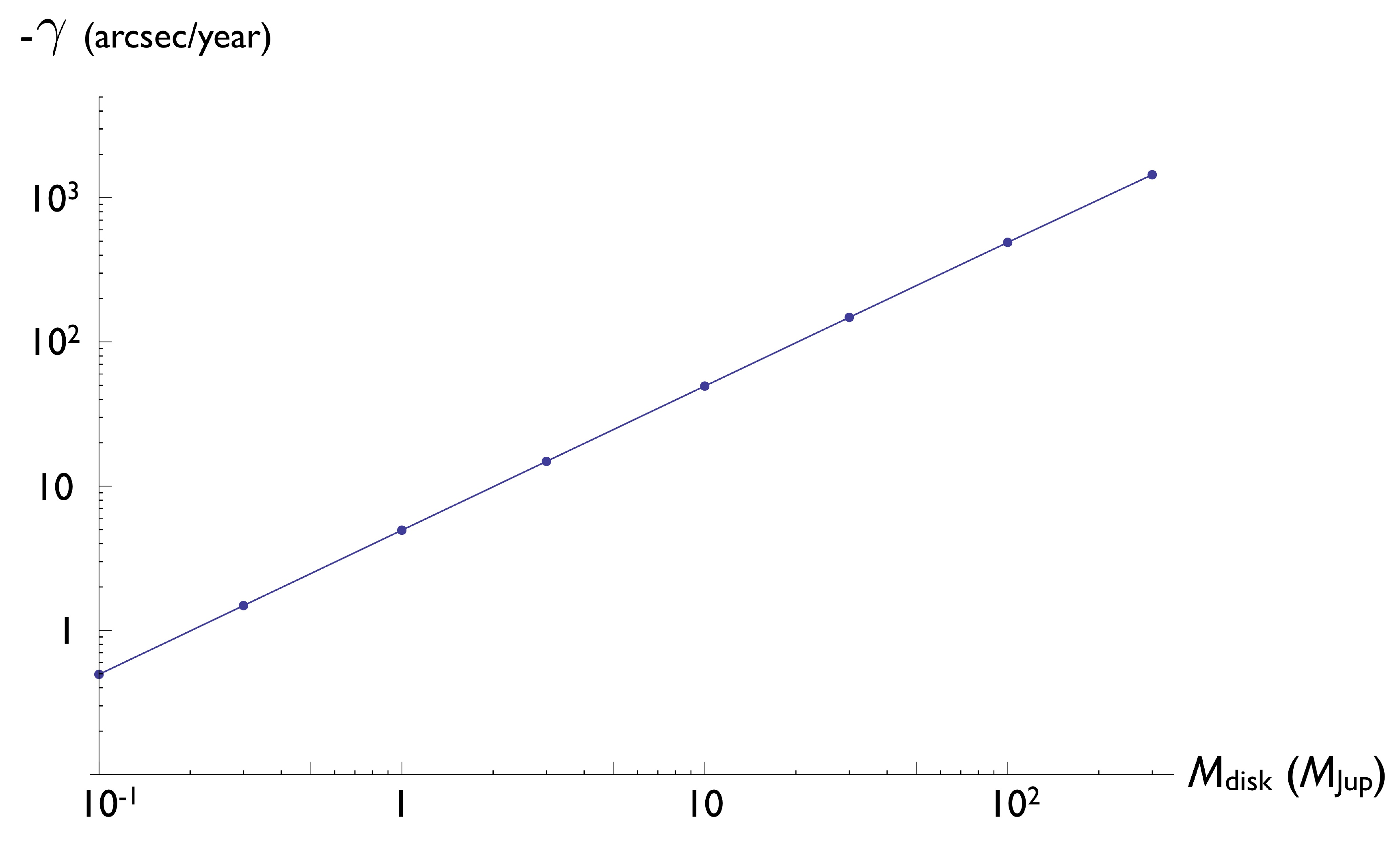}
\caption{Apsidal recession of a self-gravitating disk. The recession rate, $\gamma$ is plotted as a function of disk mass. Blue points are the model results. The points are well fit by a linear functional relationship $\gamma = -2.4 \times 10^{-5} (M_{\mathrm{disk}}/ M_{\mathrm{Jup}})$ rad/year.}
\end{figure}

Generally, typical protoplanetary disks contain $M_{\mathrm{disk}} \sim 10-100 M_{\mathrm{Jup}}$ at the time of formation, in gas and planetesimals. We have calculated the characteristic precession rate for a planetesimal embedded in such a disk, for total disk mass range, spanning roughly two orders of magnitude, between $M_{\mathrm{disk}} = 0.1 M_{\mathrm{Jup}}$ and $M_{\mathrm{disk}} = 300 M_{\mathrm{Jup}}$. Figure 2 shows the relationship between $\gamma$ and $M_{\mathrm{disk}}$. Note that unlike typical planetary systems, where secular interactions among planets give rise to positive apsidal precession, $\gamma$ of a self-gravitating disk is negative. Quantitatively, for the assumed disk geometry, the precession rate is well fit by the functional relationship $\gamma = -2.4 \times 10^{-5} (M_{\mathrm{disk}}/ M_{\mathrm{Jup}})$ rad/year. Having computed the characteristic precession rate, we can now write down the orbit-averaged Hamiltonian of a planetesimal in the disk.

We work in terms of canonically conjugated action-angle Delaunay variables
\begin{equation}
\begin{array}{lclcl}
G&=&\sqrt{a(1-e^2)} \ ,\quad \ \ \ \ \ \  g&=&\omega \nonumber \\
H&=&\sqrt{a(1-e^2)}\cos i  \ ,\quad h&=& \Omega \\
\end{array}
\end{equation}
where the inclination $i$ is measured relative to an arbitrary reference plane and  $\Omega$ is the longitude of the node of the disk relative to such a plane. In the analysis that follows, we shall take the binary star's orbital plane to be the reference plane. The Hamiltonian is simply
\begin{equation}
\mathcal{K}^{SG} = \gamma G\ .
\end{equation}
$\mathcal{K}^{SG}$ describes an eccentric precessing orbit on a fixed orbital plane (i.e. the plane of the disk).

\begin{figure*}[t]
\includegraphics[width=1\textwidth]{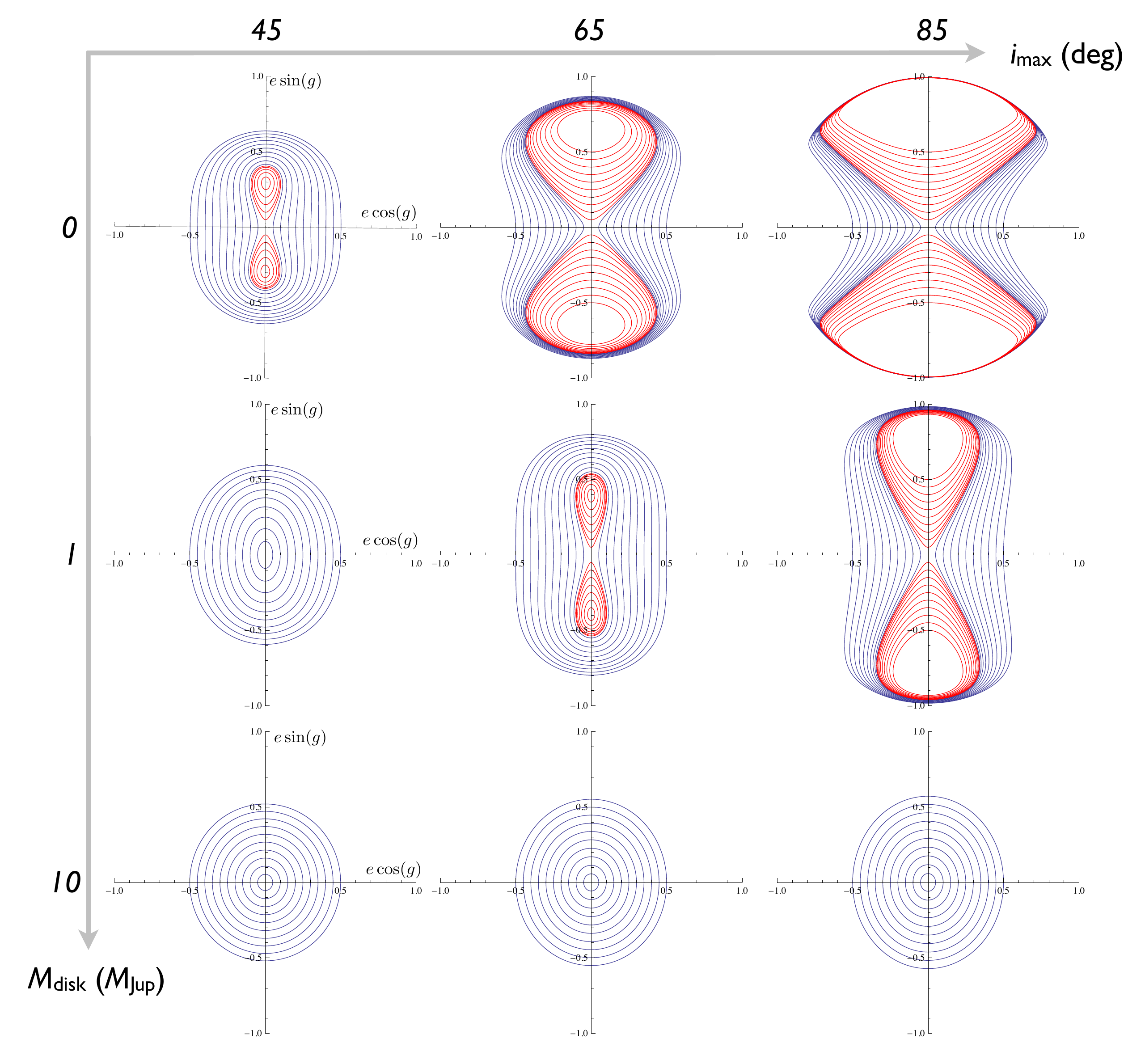}
\caption{Dynamical phase-space portraits for a planetesimal in protoplanetary disks of various masses, perturbed by a stellar companion at various inclinations showing Kozai resonance. The eccentricity vector is plotted in cartesian coordinates on each panel ($x = e \cos g, y = e \sin g$). Regions of libration of argument of perihelion are shown as red curves, while blue curves depict circulation. The top panels represent a mass-less disk, middle panels correspond to a $M_{\mathrm{disk}} = 1M_{\mathrm{Jup}}$ disk and the bottom panels show a $M_{\mathrm{disk}} = 10M_{\mathrm{Jup}}$ disk. Note that the Kozai resonance disappears as the disk mass is increased. }
\label{kozai}
\end{figure*}

Let us now incorporate the perturbations from the stellar companion into the Hamiltonian. Due to a considerable orbital separation between the protoplanetary disk and the perturber, the interactions between the two will be secular in nature. For our purposes here, we take the perturber to lie on a circular, inclined orbit. A circular perturber implies that, to leading order, potential eccentricity excitations in the disk would arise exclusively from the Kozai resonance. The free orbital precession induced by the companion can be approximated as \citep{1999ssd..book.....M}
\begin{equation}
\dot{g}_{\mathrm{free}} \simeq \frac{3 n}{2}\frac{ \tilde{m} } {M}
\left( \frac{a}{\tilde{a}} \right)^3 \end{equation} where $\tilde{m}$
and $\tilde{a}$ are the perturber's mass and semi-major axis and $M$
is the mass of the central star. In the following, we assume that
\begin{equation}
\dot{g}_{\mathrm{free}} \ll -|\gamma|\ ,
\label{condition}
\end{equation}
and neglect $\dot{g}_{\mathrm{free}}$ altogether. This condition is satisfied for well-separated binary systems ($a/\tilde{a} \ll 1$). Note that the lack of orbital eccentricity of the perturber is a mathematical convenience that makes the calculation more straight forward, without modifying our conclusions on Kozai dynamics qualitatively\footnote{An eccentric star would induce some forced eccentricity in the disk. However, because of the fast rigid apsidal precession of the disk, the forced eccentricity will be small and the disk will maintain coherence.}.

We take the stellar companion's orbital semi-major axis to be $a = 1000$AU and take the inclination as well as disk mass to be variable parameters. This choice is motivated by the estimate of the orbital separation between HD80606 and HD80607 \citep{2003ASPC..294...43E}. It is noteworthy, that the particular choice of $a$ does not have significant consequences on the dynamics of the disk, beyond setting the time-scale on which the Kozai effect operates, provided that it is large enough that condition (\ref{condition}) is satisfied. In this section, we shall assume that the nodal reference plane of the disk precesses rigidly, and the disk remains un-warped. In other words, we assume that no mutual inclination is excited between neighboring disk annuli. This feature is implicitly essential to our argument, and we will justify this assumption quantitatively in the next section.

In accord with the reasoning outlined above, we solely retain the Kozai term in the disturbing potential of the stellar companion. Consequently, the planetesimal's Hamiltonian now reads \citep{1999CeMDA..75..125K}
\begin{eqnarray}
\mathcal{K}^{SGK} &=& \gamma G + \frac{a \tilde{m} \tilde{n}^2}{(M+\tilde{m})}  \bigg\{  \frac{ 15(a - G^2) (G^2-H^2) \cos(2g)}{G^2} \nonumber \\
 &-& \frac{(5a -3 G^2) (G^2 - 3H^2)}{ G^2 } \bigg\}
\label{Ksgk}
\end{eqnarray}
where $\tilde{n}$ is the stellar perturber's mean motion. Notice that in the Hamiltonian above, $\gamma$ could be factorized, so that the magnitude of the perturbation would be proportional to $\tilde{n}/\gamma$. Since $\gamma$ is a linear function
of the disk mass, this illustrates that it is equivalent to have a closer binary companion (larger $\tilde{n}$) to a proportionally less massive disk.

With this simplified dynamical model in place, we can now explore the effect of self-gravity on the orbital excitation of the planetesimals in the disk due to the Kozai resonance. We study three choices of perturber inclination: $i = 45^{\circ}$, $i = 65^{\circ}$ and $i = 85^{\circ}$. To obtain a dynamical portrait of the system, we proceed as follows. Because the variable $h$ does not appear in (\ref{Ksgk}), the action $H$ is a constant of motion. On each $H= \mathrm{constant}$ surface,
the Hamiltonian, $\mathcal{K}^{SGK}$, describes a one-degree of freedom system, in the variables $G,g$. Simultaneously, because the Hamiltonian is also a constant of motion, the dynamics is described by the level curves of the Hamiltonian.  For simplicity, we show the dynamics in cartesian coordinates ($x = e \cos g, y = e \sin g$) in Figure \ref{kozai}, where $e$ is computed from the definition of $G$, for the assumed value of $a$ (here $a=10$AU). Given that $H$ is constant on each panel, a given eccentricity also yields the inclination. The panels are identified by the value of the inclination $i_{max}$ that corresponds to $e=0$ for the given value of $H$ (i.e. the inclination of the star relative to the initial, circular disk). Similarly, the maximal value of $e$ on each panel corresponds to $i=0$.

It is useful to begin with a discussion of a mass-less disk as this configuration is often assumed in formation studies. The corresponding plots are shown as the top panel of Figure 3. In this case, there is no added precession ($\gamma = 0$) so the Kozai resonance is present for all considered choices of inclination. The phase-space portraits show that any orbit which starts out at low eccentricity (near the origin) will follow a trajectory which will eventually lead to a highly
eccentric orbit, regardless of initial phase. In particular, for $i_{\mathrm{max}} = 45^{\circ}$, a particle which starts out on a circular orbit will attain $e_{max} \simeq 0.4$. For $i_{\mathrm{max}} = 65^{\circ}$, $e_{max} \simeq 0.85$ and for $i_{\mathrm{max}} = 85^{\circ}$, $e_{max} \simeq 1$. The resulting high-velocity collisions render formation of planetary embryos ineffective. Consequently, one should not expect planets to form under the mass-less disk approximation.

Let us now consider a $M_{\mathrm{disk}} = 1M_{\mathrm{Jup}}$ disk. The phase-space portraits of this system are shown as the middle panels of Figure 3. Although the quoted value corresponds to a very low-mass disk, the situation is considerably different from the mass-less case. For $i_{\mathrm{max}} = 45^{\circ}$, the Kozai resonance is no longer effective\footnote{Interestingly, the disappearance of the Kozai separatrix is not exactly symmetric with respect to the sign of $\gamma$. If $\gamma$ is negative, it immediately acts to erase the Kozai effect. However, a small positive $\gamma$ (i.e. $\gamma = 10^{-5}$ for $i_{max} = 45^{\circ}$) can act to enhance to Kozai effect. The effect however rapidly turns over for faster positive precession (i.e. $\gamma > 10^{-4}$).}. Thus, an initially nearly-circular orbit will retain its near-zero eccentricity, allowing for planetary formation to take place. The Kozai resonance still operates in the $i_{\mathrm{max}} = 65^{\circ}$ and $i_{\mathrm{max}} = 85^{\circ}$ cases, but the maximum eccentricities are now lower ($e_{max} \simeq 0.55$ and $e_{max} \simeq 0.95$ respectively) compared to the mass-less disk scenario.

Finally, the bottom panels of Figure 2 show the phase-space portraits of a $M_{\mathrm{disk}} = 10M_{\mathrm{Jup}}$ disk. Here, the Kozai resonance is completely wiped out, for all values of $i_{\mathrm{max}}$. Particularly, circular orbits remain circular (the center of each panel is a stable equilibrium point). Strictly speaking, these calculations describe the dynamics of planetesimals embedded in the disk. However, if planetesimals remain circular, the gaseous component of the disk must do so as well because it feels the same gravitational potential. On the other hand, if the Kozai resonance forces the planetesimals to acquire a considerable eccentricity during their evolution (the cases with low disk mass in Figure \ref{kozai} or, equivalently, cases with a close stellar companion) the gas-disk may remain more circular than the planetesimals, thanks to its additional dissipative forces. This is the situation illustrated in \cite{2011A&A...528A..40F}, where a differential evolution of planetesimals and gas, leads to size-dependent gas-drag forces.

In conclusion, recalling that $10M_{\mathrm{Jup}}$ is a lower-bound for the mass of a typical protoplanetary disk, this analysis suggests that planetary formation can take place in well separated binary systems like 16Cyg and HD80606-7 as if secular perturbations, arising from the companion star were not present.

\section{Rigid Precession of a Self-Gravitating Disk}

In the previous section, we showed that a self-gravitating disk is not succeptible to excitation by the Kozai resonance. However, in order for our argument to be complete, it remains to be shown that the assumptions of rigid precession of the disk's nodal reference plane, as well as the lack of the excitation of mutual inclination within the disk, hold true. To justify our assumptions, it is sufficient to consider a nearly circular self-gravitating disk and show that it is characterized by rigid nodal precession, since we have already shown that a flat disk will remain circular under external perturbations.

Intuitively, one can expect a rigidly precessing flat disk, from adiabatic invariance. Consider an isolated self-gravitating disk where mutual inclinations (inclinations with respect to the instantaneous mid-plane), $\hat{i}$, are initially small (i.e. $\sin \hat{i} \sim \hat{i} \ll 1$). Forced by self-gravity, the mutual inclinations within the disk will be modulated on a characteristic (secular) timescale related to the precession of the longitude of the node, $\hat{\Omega}$, relative to the disk mid-plane. One can define the action, 
\begin{equation}
J = \oint \hat{i} d \hat{\Omega} = {\rm const.} 
\end{equation}
which represents the phase-space area bounded by a secular cycle \citep{1984JApMM..48..133N}.  If the disk is subjected to an external perturbation (such as the torquing from a stellar companion), whose characteristic timescale is much longer than the secular timescale on which self-gravity modulates the mutual inclinations (the so-called adiabatic condition), $J$ will remain a conserved quantity \citep{1993PhyD...68..187H}. This implies that the mutual inclinations within the disk will remain small. This is true for each annulus of the disk in which the adiabatic condition is fulfilled. Consequently, the disk will remain unwarped and the disk mid-plane will precess rigidly at constant inclination relative to the binary star's orbital plane.

The same idea can be illustrated more quantitatively in the context of classical Laplace-Lagrange secular theory. One possible way to view the inclination dynamics of disk is by modeling the disk as a series of massive self-gravitating rings, adjacent to one-another. Note that the same technique cannot be directly applied to eccentricities, since it would predict a positive apsidal precession, whereas in reality the apsidal precession would be negative, as shown in the previous section. This is because, as soon as the eccentricity is non zero, a ring would start to intersect the adjacent rings, violating the assumption on which the Laplace-Lagrange secular theory is based.

\begin{figure}[t]
\includegraphics[width=0.5\textwidth]{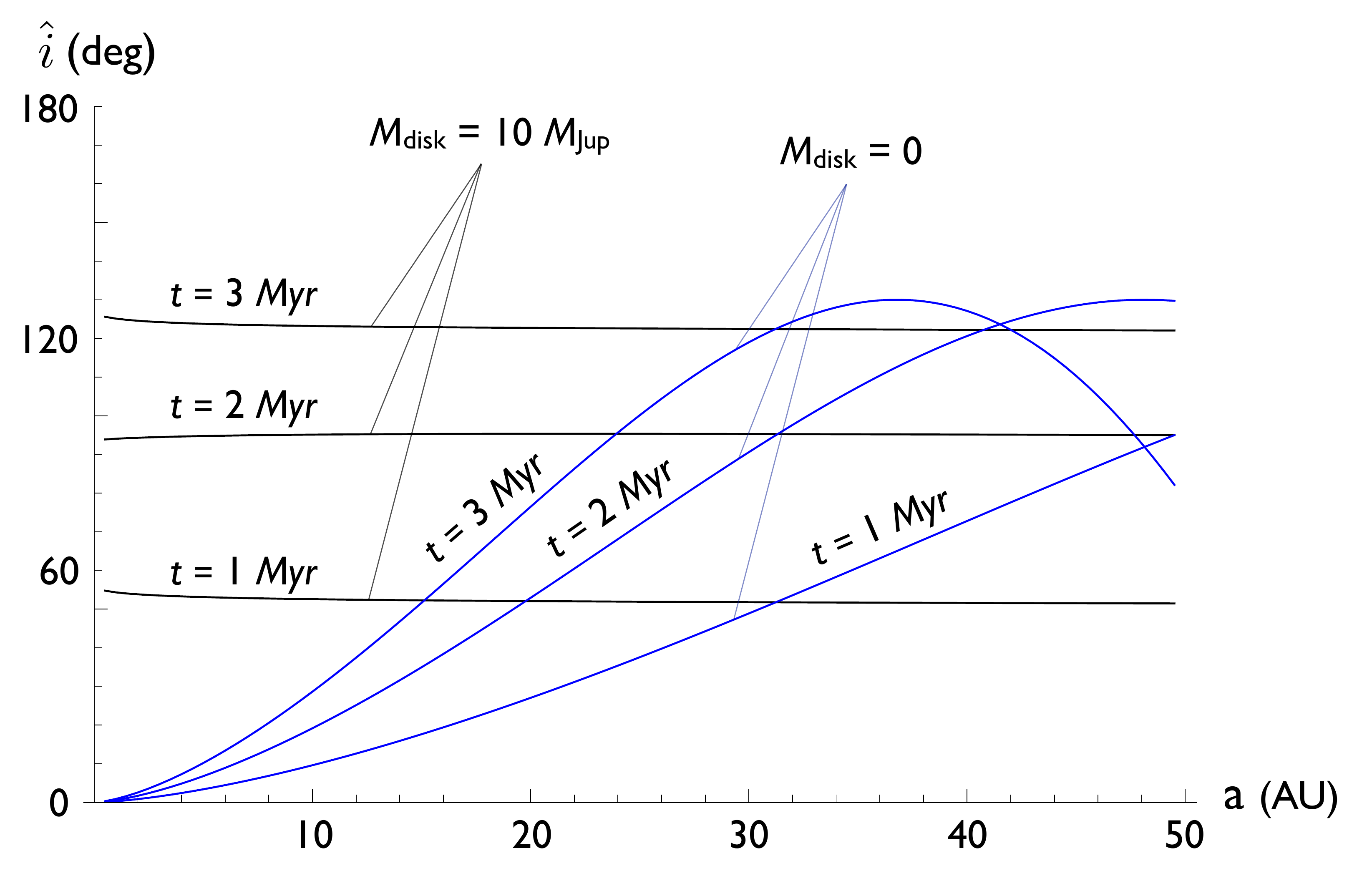}
\caption{Inclination structure of a mass-less (blue) and a self-gravitating $M_{\mathrm{disk}} = 10 M_{\mathrm{Jup}}$ disks. Here the inclination is measured relative to the original plane of the disk.  The inclination is shown as a function of semi major axis $a$ at $t = 1,3,$ and $5$ Myr. Note that the mass-less disk is considerably warped due to the perturbations from the companion star, while the self-gravitating disk maintains a uniform inclination. in this case, the growth of inclination with time is due to the rigid precession of the disk relative to the binary star plane. The inclination returns back to zero after a precession period.}
\end{figure}

The scaled Hamiltonian of a given annulus $j$, where exclusively secular terms up to second order in inclination have been retained,
reads \citep{1999ssd..book.....M}
\begin{equation}
\mathcal{K}^{LL}_j = \frac{1}{2} B_{jj} i_{j}^{'2} + 
\sum_{j=1,j\neq{k}}^{N} B_{jk} i_{j}^{'} i_{k}^{'} \cos(\Omega_{j}^{'}-\Omega_{k}^{'})
\end{equation}
where the primed quantities are expressed with respect to an fixed inertial plane (here the initial plane of the disk). In this approach, the disk is broken up into $N-1$ annuli whereas the $N^{th}$ index corresponds to the stellar companion. The coefficients $B_{jj}$ and $B_{jk}$ take the form 
\begin{eqnarray}
B_{jj} &=& -\frac{n_{j}}{4}\sum_{k=1,k\neq{j}}^{N}\frac{ {m_{k}} } {M_{\star}+m_{j}}\alpha_{jk} \bar{\alpha}_{jk} b_{3/2}^{(1)}(\alpha_{jk}) \nonumber \\
B_{jk} &=& \frac{n_{j}}{4} \frac{ {m_{k}} } {M_{\star}+m_{j}}\alpha_{jk} \bar{\alpha}_{jk} b_{3/2}^{(1)}(\alpha_{jk})
\end{eqnarray}
where $m_j$ denotes the mass of a given annulus $j$ if $j < N$, $m_N = \tilde{m}$, $\alpha = a_j/a_k $, $\bar{\alpha} = \alpha $ if perturbation is external and $\bar{\alpha} = 1 $ otherwise, while $b_{3/2}^{(1)}$ is the Laplace coefficient of the first kind. Similarly to equation (6), the diagonal terms in the $\mathbf{B}$ matrix correspond to the free nodal precession rates. To crudely account for the large mutual inclination of the stellar companion, we reduced its mass by a factor of $\sin(i)$ because in the context of second-order theory, it is appropriate to only consider the \textit{projection} of its mass onto the disk's reference plane. Rewriting the above Hamiltonian in terms of cartesian coordinates ($q = i^{'} \cos \Omega^{'}, p = i^{'} \sin \Omega^{'}$), the first-order perturbation equations ($\dot{p} = \partial \mathcal{K} / \partial{q}, \dot{q} = - \partial \mathcal{K} / \partial{p}$) yield an eigen-system that can be solved analytically (see Ch. 7 of \cite{1999ssd..book.....M}). In our calculations, we choose $N=101$ and again consider a $\sigma \propto r^{-1}$ surface density across the disk.

We take the orbital properties of the stellar companion to be the same as those discussed in the previous section and take the initial mutual inclination between the disk and the stellar companion to be $i = 65^{\circ}$. Indeed, the evaluation of the solution for various disk masses shows that the disk precesses rigidly, if the disk mass exceeds $M_{\mathrm{disk}} \gtrsim 1 M_{\mathrm{Jup}}$. This threshold is in rough quantitative agreement with the numerical models of
\cite{2011A&A...528A..40F}. Figure 4 shows the evolution of the inclination as a function of semi-major axis, of a massless (blue) disk as well as a self-gravitating (black) $M_{\mathrm{Jup}} = 10M_{\mathrm{Jup}}$ disk at various epochs. The reference plane for
the measure of the inclination is the initial plane of the disk.  We see that for the mass-less disk the inclination varies considerably with semi-major axis, which means that the disk is significantly warped, as one would expect in the context of a standard restricted 3-body problem. However, the inclination of a self-gravitating disk is nearly constant in semi-major axis, depicting an unwarped, rigid structure. Note, however, that the inclination changes with time. this is because the disk is precessing with a constant inclination relative to the plane of the binary star, so that the disk's inclination relative to its initial plane has to change periodically, over a precession period, from $i_{\mathrm{min}} = 0^{\circ}$ to a maximum of $i_{\mathrm{max}} = 130^{\circ}$ and back.

In conclusion, the assumption of untwisted structure, that we employed in the previous section when calculating the excitation by the Kozai resonance, is valid for massive disks perturbed by distant stellar companions, such as the ones that we consider in this paper. This conclusion does not apply only to the planetesimal disk, but also to the gas disk, for the same reasons mentioned at the end of the previous section.

\begin{figure}[t]
\includegraphics[width=0.5\textwidth]{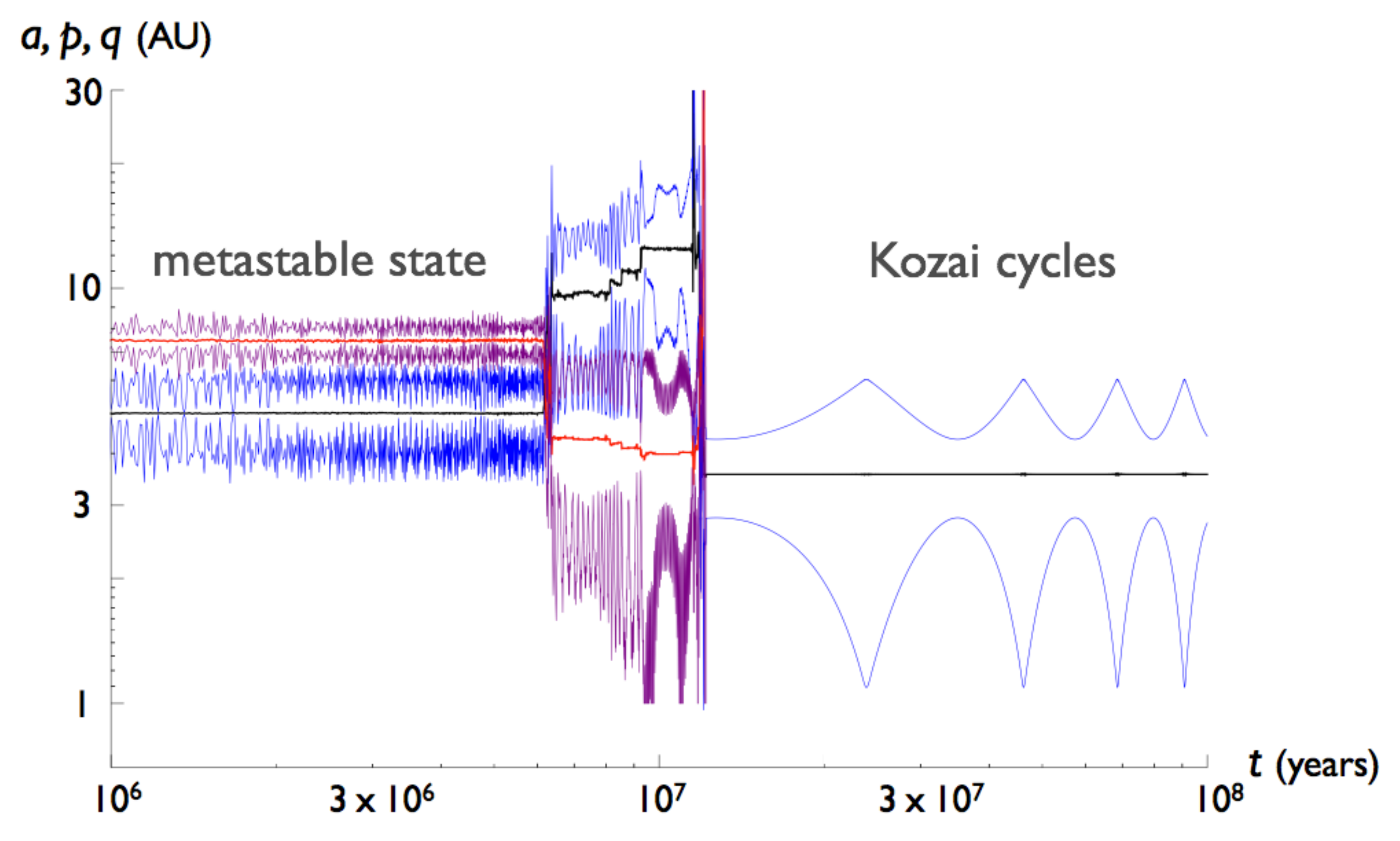}
\caption{Orbital evolution of a two-planet system and its transition into the Kozai resonance via an instability. The figure shows the semi-major axes, as well as perihelion and aphelion distances as functions of time. The planets initially start out in a metastable configuration which is protected from Kozai resonance by apsidal precession, arising from planet-planet interactions. Following $\sim 12$Myr of dynamical evolution, the planets suffer a dynamical instability, during which the initially outer planet is ejected. Consequently, the remaining planet enters the Kozai resonance.}
\end{figure}

\section{Production of Highly Eccentric Planets}

In the two preceeding sections, we have shown that in a binary stellar system, a self-gravitating disk avoids dynamical excitation, arising from the stellar companion, even if inclined. As a result, one can expect that formation of planetary systems is generally not inhibited. Furthermore, even after the disappearance of the gas, we expect that the Kozai effect can continue to be wiped out as a result of apsidal precession, induced by planet-planet interactions. As already discussed in the introduction, this is the case for the planets of the solar system with respect to the galactic tide, or the satellites of Uranus relative to solar perturbations. Consequently, the final issue we need of address is how  planets, such as HD80606b and 16Cygni Bb, do eventually end up
undergoing Kozai cycles.

The evolutionary path that a planetary system can take between the birth nebula stage and the Kozai stage is necessarily non-unique. One obvious possibility is that only a single large planet forms in the disk and as the gas evaporates, self-gravity of the disk becomes insufficient to wipe out the Kozai resonance. Such a scenario, although possible, is probably far from being universal, since the observed multiplicity in planetary systems \citep{2010HiA....15..694M}, as well as theoretical considerations \citep{2010apf..book.....A}, suggest that protoplanetary disks rarely produce only a single body.

However, an alternative picture can be envisioned: a multiple system forms and after the dispersion of the disk, still protects itself from the Kozai cycles, exerted by the companion star, through self-induced apsidal precession. Then, following a dormant period ,a dynamical instability occurs, removing all planets except one, and therefore the remaining object starts to experience the Kozai resonance. Such a scenario would be considerably more likely, since dynamical instabilities are probably common among newly-formed planetary systems \citep{2008ApJ...686..621F, 2009ApJ...699L..88R}. In particular, over the last decade or so, it has been realized that a transient dynamical instability has played an important role in shaping the architecture of the solar system \citep{1999Natur.402..635T, 2005Natur.435..459T, 2007AJ....134.1790M}. Moreover, planet-planet scattering has been suggested to be an important process in explaining the eccentricity distribution of extra-solar planets \citep{2008ApJ...686..603J} as well as the misalignment of planetary orbits with stellar rotation axes \citep{2011ApJ...729..138M}.

\begin{figure}[t]
\includegraphics[width=0.5\textwidth]{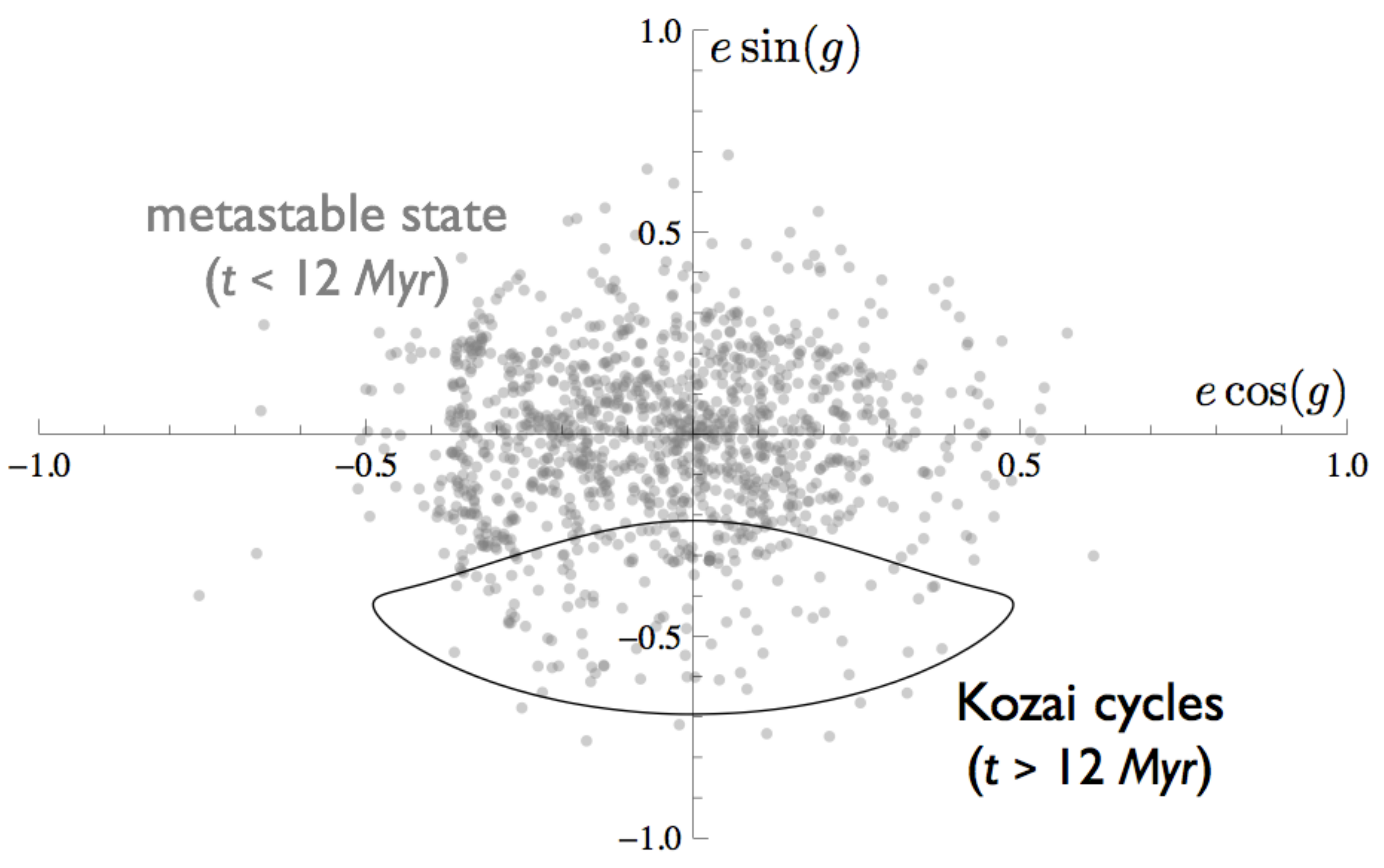}
\caption{Phase-space plot of the inner planet, corresponding to the orbital evolution, shown in Figure 4. Prior to the instability ($t<12$Myr), the motion of the planet (shown as gray points) is non-resonant. However, after a the outer planet gets ejected, the
remaining planet enters the Kozai resonance (shown as a black line).}
\end{figure}

The usefulness of analytical methods is limited when it comes to the particular study of dynamical instabilities, so one must resort to numerical methods. Below, we demonstrate a numerical proof-of-concept of the scenario outlined above.

The system we considered was a pair of giant planets, both with $m_{p} = 1 M_J$ around a sun-like ($M_{\star} = 1 M_{\odot}$) star, perturbed by a $\tilde{m} = 2 M_{\odot}$ companion. The planets were initialized on near circular orbits ($e_1 = e_2 = 0.01$) with $a_1 = 5$AU and $a_2 =7.5$AU in the same plane. The stellar companion was taken to be on a circular $\tilde{a} =1000$AU orbit, inclined by $i = 80^{\circ}$ with respect to the orbital plane of the planets. We performed the simulation using a modified version of SyMBA \citep{1998AJ....116.2067D} in which a companion star is set on a distant, fixed circular orbit. The timestep was chosen to be 0.2y and throughout the integration, the fractional energy  error remained below $\Delta E/E \leqslant 10^{-5}$. The system was evolved over $10^8$ years.

\begin{figure}[t]
\includegraphics[width=0.5\textwidth]{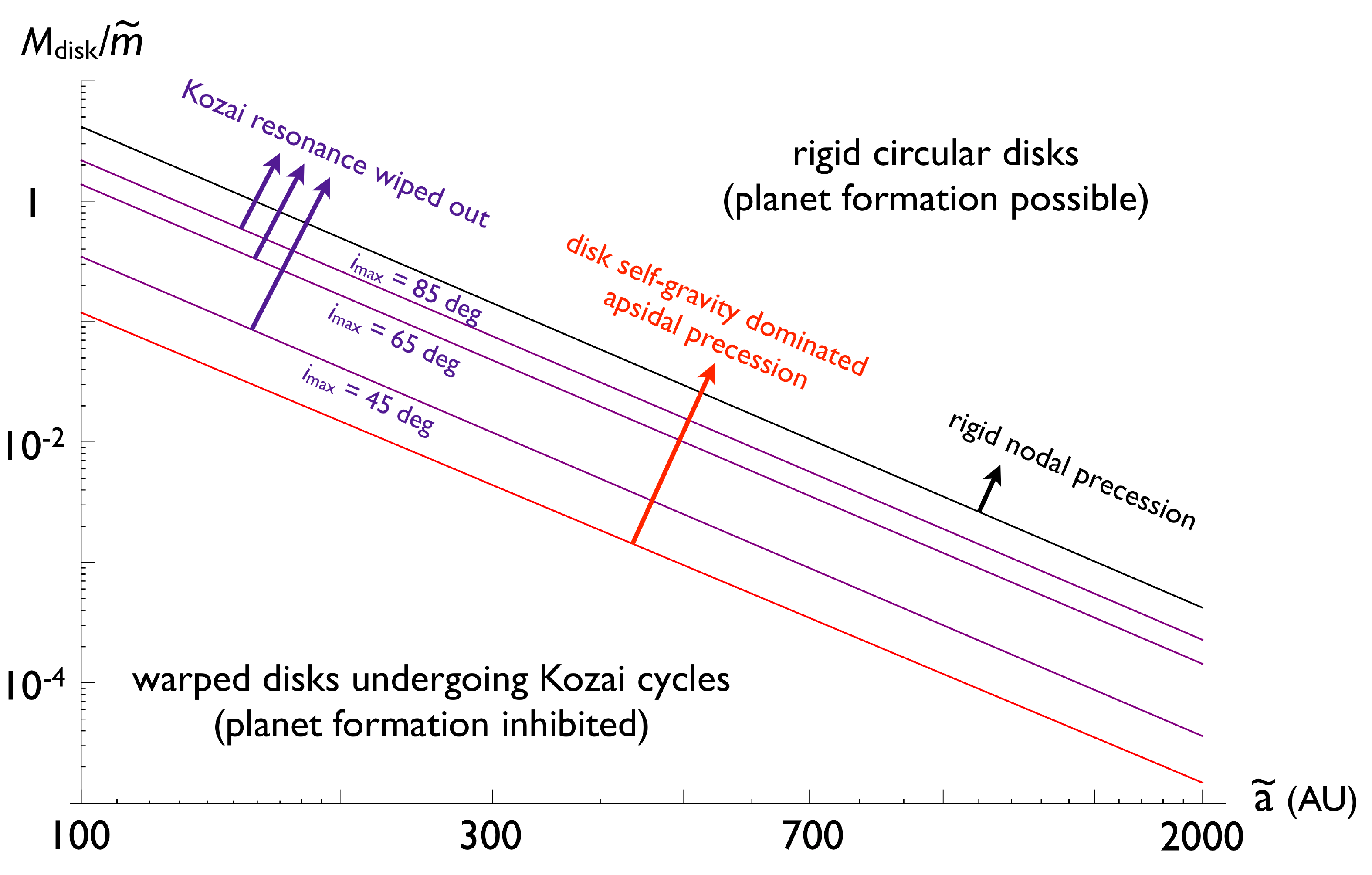}
\caption{Domain of applicability of the arguments presented in this paper. The red curve shows the dividing line between disk-dominated and stellar companion-dominated apsidal precession (as in section 2). The three purple curves illustrate the disappearance of the Kozai separatrix, for various choices of maximal inclination (as in section 3). The black curve delineates the boundary between rigid precession of the disk's mid-plane and a warped structure (as in section 4). Successful formation of planets can take place in well-separated binary systems where disk self-gravity dominates over perturbations from the stellar companion.}
\end{figure}

The orbital evolution of the system is shown in Figure 5. As can be seen, the system appears stable for the first $\sim 6$ Myr, with no sign of Kozai oscillations. However, at $\sim 6$ Myr, the system becomes unstable because the planets` orbital proximity (initial orbital separation is only $\sim 7$ Hill radii) prevents them from remaining stable on long timescales \citep{1996Icar..119..261C}.  The eccentricities of the planetary orbits grow and eventually, the planets begin to experience close encounters with each other. At $\sim12$ Myr, one of the two planets is ejected onto a hyperbolic orbit. The remaining planet, now alone and with no apsidal precession, gets captured into the Kozai resonance, and starts undergoing large, coupled oscillations in eccentricity and inclination. It is noteworthy that had the remaining planet ended up on an orbit with smaller semi-major axis, general relativistic precession could have wiped out the Kozai effect \citep{2003ApJ...589..605W, 2007ApJ...669.1298F}. Additionally, although in our setup, the stellar companion was initialized with a high inclination, this is not a necessary condition for the scenario, since the scattering event can generate mutual inclination. The transition from non-resonant motion to that characterized by Kozai cycles is depicted in Figure 6, where orbital parameters prior to the second planet's ejection are shown as gray dots and the resonant motion is shown as a black curve. Note the similarity of resonant motion computed numerically, to that computed analytically, shown in Figure 3.

\section{Discussion}

In this paper, we have addressed the issue of how planetesimals could preserve relative velocities that are slow enough to allow planet accretion to take place, in binary stellar systems. Particularly, we focused on highly inclined systems where Kozai resonance with the perturbing stellar companion have been thought to disrupt the protoplanetary disk and inhibit planet formation \citep{2009A&A...507..505M}. Here, we have shown, from analytical considerations, that fast apsidal precession, which results from the disk self-gravity, wipes out the Kozai resonance and ensures rigid precession of the disk's nodal reference plane. 

It is useful to consider the domain of applicability of the criteria discussed here. Namely, the trade-off between stellar binary separation and the perturbing companion's mass should be quantified. The region of parameter space (binary separation $\tilde{a}$ vs disk mass to perturber mass ratio) where self-gravity suppresses secular excitation from the binary companion is delineated in Figure 7. The red curve shows the dividing line between disk-dominated and stellar companion-dominated apsidal precession (as in section 2). The three purple curves illustrate the disappearance of the Kozai separatrix, for various choices of maximal inclination (as in section 3). The black curve delineates the boundary between rigid precession of the disk's mid-plane and a warped structure (as in section 4).

As can be deduced from Figure 7, for distant stellar companions ($\tilde{a} \sim 1000$ AU), the required total disk mass is of order $M_{\mathrm{disk}} \sim 1$-10$M_J$ (depending on the perturber's mass), considerably less than or comparable to, the total mass of the minimum mass solar nebula. This implies that generally, protoplanetary disks in binary stars can maintain roughly circular, unwarped and untwisted structures. Consequently, we can conclude that planetary formation in wide binary systems is qualitatively no different from planetary formation around single stars.

After the formation of planets is complete and the gaseous nebula has dissipated, the Kozai effect can continue to be inhibited as a result of orbital precession induced by planet-planet interactions. However, as the numerical experiment presented here suggests, if a planetary system experiences a transient dynamical instability that leaves the planets on sufficiently well-separated orbits, the planets can start undergoing Kozai cycles. An evolutionary sequence of this kind can explain the existence of orbital architectures characterized by highly eccentric planets, such as those of HD 80606 and 16 Cygni B \citep{2001ApJ...562.1012E, 2003ApJ...589..605W}.

The work presented here resolves, at least in part, a pressing dynamical issue of planetary formation in highly inclined binary systems. As an avenue for further studies, the analytical results presented here should be explored numerically in grater
detail. Particularly, hydrodynamic simulations, such as those presented by \cite{2011A&A...528A..40F} can be used to quantitatively map out the parameter space that allows for planetary systems to form successfully.

The study presented here has further consequences beyond an explanation of planet formation in wide binary systems. Particularly, the model of instability-driven evolution of newly-formed systems into the Kozai resonance has substantial implications for orbital misalignment with the parent star's rotation axis. In fact, Kozai cycles with tidal friction produce a particular distribution of orbit-spin axis angles \citep{2007ApJ...669.1298F}. This distribution differs significantly from that produced by the planet-planet
scattering scenario \citep{2008ApJ...678..498N}. This distinction has been used to statistically infer the dominant process by which misaligned hot Jupiters form \citep{2011ApJ...729..138M}. However, the model presented here suggests that the two distributions should be intimately related, as planet-planet scattering provides the initial condition from which Kozai cycles originate. Consequently, a quantitative re-examination of the orbit-spin axis misalignment angle distribution, formed by Kozai cycles with tidal friction that originate from a scattered orbital architecture, and subsequent comparison of the results with observations of the Rossiter-McLaughlin effect will likely yield new insights into dynamical evolution histories of misaligned hot Jupiters.

\textbf{Acknowledgments} We thank the referee, Y. Wu for useful suggestions.

{}

\end{document}